\RequirePackage{fix-cm}
\documentclass[twocolumn,epjc3]{svjour3}
\smartqed  
\RequirePackage{graphicx} \DeclareGraphicsExtensions{.eps,.ps}
\usepackage[latin1]{inputenc}
\usepackage{epstopdf}
\RequirePackage{amssymb}
\journalname{Eur. Phys. J. C}
\begin{document}

\title{The photon magnetic moment problem revisited
} 


\author{ H. Pérez Rojas\thanksref{e1,addr1}
        \and
        E. Rodríguez Querts\thanksref{e2,addr1,addr2} 
}

\thankstext{e1}{e-mail: hugo@icimaf.cu}
\thankstext{e2}{e-mail: elizabeth@icimaf.cu}

\institute{Departamento de Física Teórica, Instituto de Cibernética,
Matemática y Física, Calle E 309, Vedado, Ciudad Habana, Cuba.
\label{addr1}
                     \and
           \emph{Present Address:}  ICTP, Strada Costiera 11, 34014 Trieste, Italy \label{addr2}
}

\date{Received: date / Accepted: date}

\maketitle

\begin{abstract}
The photon magnetic moment for radiation propagating in magnetized
vacuum is defined as a pseudo-tensor quantity, proportional to the
external electromagnetic field tensor. After expanding the
eigenvalues of the polarization operator in powers of $k^2$, we
obtain  approximate dispersion equations (cubic in $k^2$), and
analytic solutions for the photon magnetic moment, valid for low
momentum and/or large magnetic field.
 The paramagnetic photon experiences a red
shift, with opposite sign than the gravitational one, which differs
for parallel and perpendicular polarizations. It is due to the drain
of photon transverse momentum and energy by the external field. By
defining an effective transverse momentum, the constancy of the
speed of light orthogonal to the field is guaranteed. We conclude
that the propagation of the photon non-parallel to the magnetic
behaves as if there is a quantum compression of vacuum or warp of
space-time in an amount depending on its angle with regard to the
field. 
\end{abstract}
\section{Introduction}
We have shown in \cite{EliPRD} that for a photon moving in a
magnetic field $\mathbf{B}$, assumed constant and homogeneous (and
for definiteness, taken along the $x_3$ axis, thus $|B_3|=B$,
$B_1=B_2=0$ \footnote{Our statements below are valid for all frames
of reference moving parallel to $\mathbf{B}$.}), an anomalous
magnetic moment defined as $\mu_{\gamma}=-\partial \omega/\partial
B$ arises. This quantity \textit{has meaning, and can be defined
only} when the photon mass shell includes the radiative corrections,
i.e., the magnetized photon self-energy, and calculated explicitly
only \textit{after} obtaining the solution of the photon dispersion
equations \cite{shabad2}. It was shown that it is paramagnetic
($\mu_{\gamma}>0$), since it arises physically when the photon
propagates, as due to the magnetic response of the virtual
electron-positron pairs of vacuum, or vacuum polarization, under the
action of $\mathbf{B}$, leading to vacuum magnetization. Thus, the
photon embodies both properties of the free photon and of a magnetic
dipole, which leads to consider it more as a quasi-photon, in
analogy with the polariton of condensed matter physics
\cite{polariton}. Such properties are valid in the whole region of
transparency, which is the region of momentum space where the photon
self-energy, and in consequence, its frequency $\omega$, is real.
This region is defined for transverse momentum
$(\omega^2-k_{\parallel}^2)^{1/2}\leq 2m$, where $\omega$,
$k_{\parallel}$,  are the photon frequency and momentum components
along $\mathbf{B}$, and $m$ is the electron mass. In \cite{shabad2}
it is shown that the quantities
\begin{eqnarray}\label{1}
z_1&=&(\mathbf{k} \cdot
\mathbf{B})^2/\mathbf{B}^2-\omega^2=k_\parallel^2-\omega^2,\\
\nonumber z_2&=&(\mathbf{B} \times {\mathbf{k}})^2/\mathbf{B}^2\\
\nonumber &=&\mathbf{k}^2-(\mathbf{k} \cdot
\mathbf{B})^2/\mathbf{B}^2=k_\perp^2,
\end{eqnarray}
are relativistic invariant variables for the photon propagating in
the magnetic field $\mathbf{B}$, where $z_1 + z_2 =k^2$ is the
square of the four-momentum vector $k_\mu$. In what follows we will
use equally $z_2$ and $k_{\perp}^2$ when referring to the transverse
momentum squared.

As pointed out in \cite{EliPRD}, beyond that region, as the photon
becomes unstable \cite{shabad2} for frequencies $\omega\geq 2m$ (and
has a significant probability of decaying in electron-positron
pairs), the photon magnetic moment loses meaning if considered
independent of the magnetic moment produced by the electron-positron
background. Let us remark that  the case studied in \cite{EliPRD},
\cite{shabad2} is based on the hypothesis of a constant and
homogeneous magnetic field defined by the invariants
$\textsf{F}={\cal F}_{\mu \nu}^2=2(B^2-E^2)=const>0$, $\textsf{G}
=\mathbf{E}\cdot \mathbf{B}=0$ (as pointed out earlier, we will
refer to the set of frames for which the external field $\bf{E}=0$).
Expressions for physical quantities as the polarization operator
$\Pi_{\mu \nu}$ depend on scalar quantities such as $\textsf{F} =
2B^2$, $k^2$ (the total four-momentum squared) and $k_\mu {\cal
F_{\mu \nu}}^2 k_\nu$. Being scalars, they do not depend on the
direction of the coordinate axis, although in a specific problem a
direction for $\mathbf{B}$ must be chosen. Such a direction breaks
the spatial symmetry, and for simplicity, it is chosen as coinciding
with one of the coordinate axes. In \cite{EliPRD} we found the
expressions of the photon magnetic moment keeping in mind that the
exact photon self-energy is an even function of $B$, as it is
demanded by Furry's theorem \cite{Fradkin}).

Also in \cite{EliPRD} the  dynamical results obtained have general
validity in the subset of Lorentz frames moving parallel to the
magnetic field pseudovector $\mathbf{B}$, independently of the
orientation of the coordinate axes, since 3D scalars (as
$\mathbf{k}^2$) and pseudoscalars (as $\mathbf{B}\cdot \mathbf{k}$)
are invariant under proper rotations. The reduced Lorentz symmetry
for a specified chosen field direction $\mathbf{B}$ is obviously
described by the group of Lorentz translations along $\mathbf{B}$
multiplied by the group of spatial rotations around $\mathbf{B}$. In
\cite{Selym}, \cite{SelymShabad} a ``perpendicular component" for
the photon magnetic moment, orthogonal to $\mathbf{B}$, was reported
to exist as a non-zero vector, but as was recognized by the authors,
it does not contribute to the photon energy and can not be deduced
from the
photon dispersion equation, 
as we will show in Section 2.

In Section 2 it is shown in a neat way that the photon magnetic
moment introduced in \cite{EliPRD} can be defined as a quantity
linear in the external electromagnetic field tensor ${\cal{F}_{\mu
\nu}}$, from which a pseudovector photon magnetic moment
${\mathbf{\mu}}_{\gamma}^{(i)}\parallel \mathbf{B}$ can be written
for each Lorentz frame parallel to $\textbf{B}$. Physical reasons
are given later in support to this fact.  We focus mainly in the
case in which the dispersion law is a small deviation of the light
cone case. We shall introduce a cubic in $k^2$ approximation for the
dispersion curve, whose region of validity cover most of the region
of transparency for fields very near the critical $B \lesssim B_c$,
(where $B_c =m^2/e \sim 4.14 \times 10^{13}$ G is the Schwinger
critical field) but decreases for supercritical fields, and they can
be compared to the exact curves, drawn numerically for fields of
order and greater than $B_c$. We conclude that in the region of
transparency the paramagnetic photon behavior is maintained for
supercritical fields. In Section 3 we discuss the interesting
consequence of photon paramagnetism, which leads to a decrease of
the frequency with increasing magnetic field. The effect is
polarization-dependent. We interpret that in such region the speed
of light does not change, but the dispersion law must be
reinterpreted by defining an effective transverse momentum which
decreases with $B$. This leads to space-time consequences: vacuum
orthogonal to the field behaves as compressed; time, measured by the
period of an electromagnetic wave, run faster for increasing $B$ and
is direction-dependent.

In Section 4 we deal in a more detailed way with the red shift
effect in a magnetic field (already reported in
\cite{ElizabethPROC}), which acts in an opposite way than the
gravitational red shift in the whole range of the transparency
region. We discuss also the arising of an effective transverse
momentum orthogonal to the field (also polarization-dependent), from
which the photon dispersion curves are obtained in a wide range of
frequencies characterized by the condition $z_1 <<m^2$.

\section{Photon magnetic moment
from tensor and pseudovector expressions} In \cite{EliPRD}  the
quantity $\partial \omega/\partial B$ was introduced as the modulus
of a vector parallel to $\mathbf{B}$. The definition of photon
magnetic moment was a generalization of the usual definition of this
quantity for electrons and positrons, as is done in \cite{Johnson}.
Then $\mu_{\gamma}=-\partial \omega/\partial B$ is understood as the
modulus of a vector along $\mathbf{B}$ since  we have $\partial
B/\partial \mathbf{B}=\mathbf{n}_{\parallel}$, where
$\mathbf{n}_{\parallel}$ is a unit vector parallel to $\mathbf{B}$.

Let us consider the expression for the vector
${\mathbf{\mu}}_{\gamma}=-\partial \omega/\partial \mathbf{B}$ in
the most general case. For any value of $B$ and independently of the
order considered in the loop expansion for the polarization
operator, the photon anomalous magnetic moment will be shown to be a
vector parallel to $\mathbf{B}$. This can be easily deduced from the
photon dispersion equations. Initially we have seven independent
variables: the four components of $k_\mu$ plus the three components
of $\mathbf{B}$ in an arbitrary system of reference. By choosing the
field along a fixed axis, say, $x_3$, its three components are
reduced to one $B= \sqrt{\textsf{F}/2}$ (in components it is
$B_{\mu}= \frac{1}{2}\varepsilon_{\mu \lambda \nu}{\cal F}_{\lambda
\nu}$).
 Each of the dispersion equations for the eigenvalues of the
polarization operator $\kappa^{(i)}$ ($i=1,2,3$) imposes an
additional constraint, reducing the independent variables to four,
$B$ plus the three components of $\mathbf{k}$ which are $k_1, k_2$
and $k_3\equiv k_{\parallel}$, but cylindrical symmetry around
$\bf{B}$ makes  $k_1,k_2$ to appear always as $k_1^2 + k_2^2=z_2$,
reducing one independent variable. As $\kappa^{(i)}$ depends on the
photon momentum components in terms of the invariant variables $z_1,
z_2$, the dispersion equations, obtained as the zeros of the photon
inverse Green function $D^{-1}_{\mu\nu}=0$, after diagonalizing the
polarization operator, are

\begin{equation}
k^2=\kappa_{i}(z_2,z_1,B) \hspace{1cm} i=1,2,3.
\end{equation}
which can be written \cite{shabad2} as
\begin{equation}
  z_1 + z_2=\kappa^{(i)}(z_1,z_2,B),\hspace{1cm}i=1,2,3.
\label{dispeq}
\end{equation}
There are three non vanishing eigenvalues and three eigenvectors,
since $i=1,2,3$, corresponding to three photon propagation modes.
One additional eigenvector is the photon four momentum vector
$k_{\nu}$ whose eigenvalue is $\kappa_{4}=0$. \cite{shabad2}.
However, in a specific direction only two, out of the three modes,
propagate in vacuum, which manifests the property of bi-refringence.

The independent variables in (\ref{dispeq}) are reduced to two, for
instance, $z_2$ and $B$, if (\ref{dispeq}) is solved as $z_1=f(z_2,
B)$ \cite{shabad2}. But as $k_{\parallel}$ is a component of the
photon momentum, the dependence of $z_1$ on $z_2$ and $B$ in
specific calculations is assumed as being contained on the photon
energy $\omega$. Thus we usually write $\omega^2=
k_{\parallel}^2-f^{(i)}(z_2, B)$. In other words, in the solution of
each of the dispersion equations one assumes $\omega^2$ as a
function of the independent variables $z_2, k_{\parallel}$ and $B$.
After solving the dispersion equations for $\omega$ in terms of
$k_{\parallel}$ and $z_2$ we get
\begin{equation}
\omega^{(i)2}=\vert\textbf{k}\vert^2+ f\left(z_2,m^2,
B\right)^{(i)}. \label{eg2}
\end{equation}
It can be shown \cite{shabad2} that
 for propagation  orthogonal to $B$ the mode $i=2$ is polarized along
$B$ and the $i=3$ is polarized perpendicular to $B$.

We will define from (\ref{eg2}) the tensor
\begin{eqnarray}
M_{\mu \nu}^{(i)}&=&\partial z_1/\partial \cal{F}_{\mu
\nu}\\
\nonumber &=&\frac{1}{2}\frac{\partial f^{i}(z_2,B)}{\partial
B^2}\cal{F}_{\mu \nu},
\end{eqnarray}
Thus, \begin{equation}-\frac{\partial \omega}{\partial {\cal{F}_{\mu
\nu}}}=\frac{1}{2\omega}\frac{\partial z_1}{\partial {\cal{F}_{\mu
\nu}}}= \frac{1}{2\omega} M_{\mu
\nu}^{(i)}=\frac{1}{4\omega}\frac{\partial z_1}{\partial B^2}
{\cal{F}_{\mu \nu}}.\end{equation} Then the photon magnetic moment
can be defined as a quantity proportional to the pseudovector
 \begin{equation}M_{\lambda}=\frac{1}{2}\frac{\partial z_1}{\partial
 B^2}
\varepsilon_{\lambda \mu \nu } {\cal{F}_{\mu \nu}}.\end{equation}
The proportionality factor $1/2\omega$ is not Lorentz invariant, and
for each frame moving parallel to $\mathbf{B}$ we define for each
mode the photon magnetic moment as a $3d$ pseudo-vector
${\mathbf{\mu}}_{\gamma}^{(i)}=\frac{1}{2\omega}\mathbf{M}$, where
$\mathbf{M}||\mathbf{B}$.

This can also be seen directly from (\ref{dispeq}) by writing
\begin{equation}\label{der1}
\frac{\partial z_1}{\partial B} =\frac{\partial
\kappa^{(i)}}{\partial z_1}\frac{\partial z_1}{\partial
B}+\frac{\partial \kappa^{(i)}}{\partial B},
\end{equation}
which leads to
\begin{equation}\label{der2}
\frac{\partial z_1}{\partial B}=-2\omega
\frac{\partial\omega}{\partial B} =\frac{\frac{\partial
\kappa^{(i)}}{\partial B}}{1-\frac{\partial \kappa^{(i)}}{\partial
z_1}}.
\end{equation}
Finally we get  the $3d$ pseudo-vector photon anomalous magnetic
moment as
\begin{eqnarray}\label{magneticmom}
   {\mathbf{\mu}}_{\gamma}^{(i)}& \equiv & -\frac{\partial \omega}{\partial
   \mathbf{B}}\\
   &=&\frac{1}{2\omega}\frac{\frac{\partial \kappa^{(i)}}{\partial B}}
   {1-\frac{\partial \kappa^{(i)}}{\partial
   z_1}}\mathbf{n}_{\parallel}.\nonumber
\end{eqnarray}
Thus, $|M|=\frac{\partial \kappa^{(i)}}{\partial B}/
   (1-\frac{\partial \kappa^{(i)}}{\partial
   z_1})$. It has been proved in the most general case that
${\mathbf{\mu}}_{\gamma}=-\partial \omega/\partial
   \mathbf{B}=-(\partial \omega/\partial B)\mathbf{n}_{\parallel}$
   is a vector parallel to $\mathbf{B}$.

\subsection{Conserved electron-positron and photon angular momentum}
For the transparency region, ($\omega<2m$) the photon magnetic
moment is a consequence of the vacuum magnetization produced by the
electron-positron virtual pairs. The dynamics of observable
electrons and positrons was discussed  in \cite{Johnson}, and these
results are valid for virtual pairs of vacuum. All symmetry and
conservation properties are valid for vacuum pairs, in agreement to
the content of a basic theorem due to Coleman \cite{Coleman} which
states that \textit{the invariance of the vacuum is the invariance
of the world}.

As stated earlier, for electrons and positrons physical quantities
are invariant only under rotations around $x_3$ or displacements
along it \cite{Johnson}. This means that conserved quantities (whose
operators commute with the Dirac Hamiltonian), are all parallel to
$\mathbf{B}$, as angular momentum and spin components
$\mathbf{J}_3$,$\mathbf{L}_3$,$\mathbf{s}_3$ and the linear momentum
$\mathbf{p}_3$. We must emphasize here that \textit{the
electron-positron momentum orthogonal to $\textbf{B}$ is not
conserved}. \textit{It implies that for the photon dispersion
equation, which includes the self-energy tensor, momentum
$k_{\perp}$ orthogonal to the field is neither conserved}. Also,
eigenvalues $J_{1,2}$, $L_{1,2}$, $s_{1,2}$ do not correspond to any
observable. By using units $\hbar=c=1$, the energy eigenvalues are
$E_{n,p_3}=\sqrt{p_3^2+m^2+ eB(2n+1+s_3)}$ where $s_3=\pm 1$ are the
spin eigenvalues along $x_3$ and $n=0,1,2..$ are the Landau quantum
numbers. In other words, the transverse squared Hamiltonian
$\mathbf{H}_t^2$ eigenvalues are
$E_{n,p_3}^2-p_3^2-m^2=eB(2n+1+s_3)$, it is quantized as integer
multiples of $eB$. It can be written
$\mathbf{H}^2_t=2eB(\mathbf{J}_z + eB \mathbf{r}_0^2/2)$, where
$\mathbf{r}_0^2$ is the squared center of the orbit coordinates
operator, with eigenvalues $(2l+1)/eB$, and the eigenvalues of $J_z$
are $n-l + s_3/2$. Thus, the energy is degenerate with regard to the
quantum number $l$, or either, with regard to the momentum $p_y$ or
the orbit's center coordinate $x_0=p_y/eB$.

The magnetic moment operator $\mathbf{M}$ is the sum of two terms
one of which \cite{Johnson} is not a constant of motion but its
\textit{quantum average} vanishes. Its expectation value is $\bar
M=-<\partial \mathbf{H}/\partial B>=-\partial E_{n,p_3}/\partial B$
\cite{Pauli}. Then
\begin{equation}
\bar M(p_3,n)=-(E^2-p_3^2-m^2)/2B E \label{epMM},
\end{equation}  is the modulus of a vector parallel to $\mathbf{B}$ for negative
energy states, antiparallel to $\mathbf{B}$ for positive energy
states, and $\mathbf{B}= M\mathbf{n}_{\parallel}$.

The expression (\ref{epMM}) for the magnetic moment behaves as
\textit{diamagnetic}, but the magnetization, obtained from the
energy density of vacuum has otherwise a \textit{paramagnetic}
behavior. This is  because, due to the degeneracy of energy
eigenvalues with regard to the orbit's center coordinates, the
density of states depends linearly on the magnetic field (returning
momentarily to units $\hbar, c$) through the factor
$eB/4\pi^2\hbar^2c^2$. Thus (\ref{epMM}) is not enough for
calculating the vacuum magnetization since we must start actually
from the energy eigenvalues, and taking the density of states
factor, proceed to sum over Landau states $\sum_n$ and to integrate
on $\int c dp_3$. Then, for obtaining the energy density, we must
note that the factor $(1/\hbar c)\int c dp_3 $ provides energy per
unit length whereas the factor $(eB/\hbar c)$, having inverse square
of length dimensions and
 coming from the
orbit's center degeneracy, is necessary to provide energy per unit
volume. By recalling that $\phi_0=\hbar c/e$ is the magnetic flux
quantum, the term in parenthesis can be written as $B/\phi_0$ and it
is (up to a factor $1/4\pi^2$) the number of flux quanta per unit
area orthogonal to the field in vacuum. Thus, we see that due to
this factor the Landau ground state $n=0$, whose energy eigenvalue
is independent of $B$, has however, an important contribution to
vacuum magnetization.

Notice that, although $<\mathbf{M}>$ and $\mathbf{J}_3$ are parallel
vectors, and these quantities are closely related dynamically, there
is no a linear relation between their moduli $\bar M$ and $J_3$ as
it is in non-relativistic quantum mechanics. On the opposite $\bar
M$ is a nonlinear function of the $J_3$ and and $r_0^2$ eigenvalues.
There is no room for an electron magnetic moment component
orthogonal to $\mathbf{B}$, which would provide a physical basis to
that of the photons.

To obtain the expression for the vacuum energy density $\Omega$ we
start  from
\begin{equation}\Omega
=(eB/4\pi^2\hbar^2 c)\sum_n \int dp_3  \alpha_n E(p_3, n,
eB),\end{equation} where $\alpha_n=2- \delta_{0n}$,
 is a degeneracy factor.
Such expression  is divergent, and after subtracting the
divergences, one is left with the Euler-Heisenberg expression for
the vacuum energy (returning to units $\hbar =c=1$),
$\Omega_{EH}=\frac{\alpha B^2}{8\pi^2}\int_0^{\infty}e^{-B_c
x/B}\left[\frac{coth x}{x} -\frac{1}{x^2}-\frac{1}{3}\right]\frac{d
x}{x}$ which is an even function of $B$ and $B_c$,where
$B_c=m^2/e\simeq 4.4\times 10^{13}$G is the Schwinger critical
field. The main conclusion is that magnetized vacuum is paramagnetic
${\cal M}_V=-\partial \Omega_{EH}/\partial B>0$ and is an odd
function of $B$ \cite{Elizabeth}. For $B<<B_c$ it is ${\cal
M}_V=\frac{2\alpha}{45 \pi}\frac{B^{3}}{B_c^2}$, where $\alpha$ is
the fine structure constant.
\begin{figure}
\centering
\includegraphics[width=0.60\textwidth]{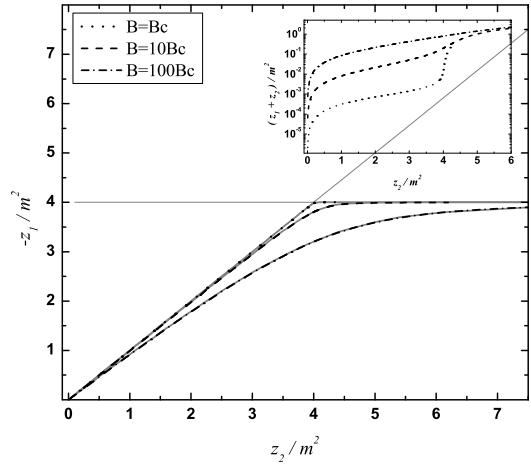}
\caption{ Solutions of the dispersion equations for the second mode,
for different magnetic field values (with continuous lines we
represent the approximate solution and with discontinuous lines the
exact ones). Note that the light cone curve is the straight line
$-z_1=z_2$. The behavior for $(z_1+z_2)/m^2$ is drawn in the upper
right figure, in a logarithmic scale, to allow depict the three
curves. } \label{fig:mb2}
\end{figure}
We conclude that no  component of ${\mathbf{\mu}}_{\gamma}$
perpendicular to $\mathbf{B}$ arises in our problem. But the main
conclusion according to  \cite{EliPRD}, after explicit calculations,
is the photon paramagnetic behavior ${\mathbf{\mu}}_{\gamma}>0$.

\section{Photon anomalous magnetic moment in the one loop approximation}
We want to give explicit expressions for the photon magnetic moment,
starting from the renormalized eigenvalues of the polarization
operator in the one-loop approximation, given in \cite{shabad2}
\begin{equation}\label{op-pol2}
\kappa_{i}=\frac{2 \alpha}{\pi}\int_{0}^{\infty}dt
\int_{-1}^{1}d\eta e^{-\frac{t}{b} }\left[\frac{\rho_{i}}{\sinh t}
e^{\zeta}+\frac{k^2 \bar{\eta}^2}{2t} \right],
\end{equation}
\begin{eqnarray*}
\zeta(z_1,z_2,B)&=&-\frac{z_{2}}{eB}\frac{\sinh (\eta_+ t)\sinh
(\eta_-
t)}{\sinh t}-\frac{z_1 }{eB}\bar{\eta}^2t,\\
 \rho_1(z_1,z_2)&=&-\frac{k^2}{2}\frac{\sinh (\eta_+ t)\cosh (\eta_+t)}{\sinh
 t}\eta_-,
\\
 \rho_2(z_1,z_2)&=&-\frac{z_1}{2}\bar{\eta}^2 \cosh t-\frac{z_2}{2}
 \frac{\sinh (\eta_+ t)\cosh (\eta_+t)}{\sinh t}\eta_-,
\\
 \rho_3(z_1,z_2)&=&-\frac{z_2}{2}\frac{\sinh (\eta_+ t)\sinh (\eta_- t)}
 {\sinh^{2} t}\\& &\hspace{2cm} -\frac{z_1}{2}\frac{\sinh (\eta_+ t)\cosh (\eta_+
t)}{\sinh t} \eta_-,
\end{eqnarray*}
where we have used the notation $b=\frac{eB}{m^2}=\frac{B}{B_c}$,
$\eta_{\pm}=\frac{1 \pm \eta}{2}$, $\bar{\eta}=\sqrt{\eta_+\eta_-}$.

As we  discussed in \cite{EliPRD}, an explicit expression for a
photon magnetic moment $\mu_{\gamma}^{2,3}>0$ in the regions
$-z_1\leq 4m^2$  can be obtained from (\ref{op-pol2}). To that end
we differentiate with regard to $B$ the dispersion equation
$z_{1}+z_{2}=\kappa_{i}$ and get
\begin{equation} \label{FAMM1} \nonumber
  \frac{ \partial z_{1}}{\partial B} = \frac{\partial \kappa_{i}}{\partial
  B} =\frac{2 \alpha}{\pi}\int_{0}^{\infty}dt
\int_{-1}^{1}d\eta e^{-\frac{t}{b}}\left[ \phi_{i} +\frac{\partial
z_{1}}{\partial B}\varphi_{i }\right],
\end{equation}
\begin{eqnarray*}
\phi_{i}&=& \frac{1}{m^2}\left[\frac{\rho_{i}e^{\zeta}}{\sinh t}
\left(\frac{t}{b} -\zeta\right)+ \frac{k^2}{b}\frac{\bar{\eta}
^2}{2}\right] ,
\\  \varphi_{i }&=&
\frac{e^{\zeta}}{\sinh t} \left(\frac{\partial \rho_{i}}{\partial
z_{1}}-\frac{ \rho_{i}}{eB}\bar{\eta }^2t \right)+\frac{\bar{\eta
}^2}{2t},
\end{eqnarray*}
and, taking in mind that $\frac{\partial z_{1}}{\partial B}=-2\omega
\frac{\partial \omega}{\partial B}$ in (\ref{FAMM1}), we obtain a
general expression for the photon anomalous magnetic moment
\begin{equation}\label{FAMM2} \nonumber
   \mu_{\gamma}^i=-\frac{\partial \omega}{\partial B} = \frac{m^2}{2 \omega B}\frac{\frac{2 \alpha}{\pi}\int_{0}^{\infty}dt
\int_{-1}^{1}d\eta e^{-\frac{t}{b} } \phi_{i}}{1-\frac{2
\alpha}{\pi}\int_{0}^{\infty}dt \int_{-1}^{1}d\eta e^{-\frac{t }{b}}
\varphi_{i }}.
\end{equation}

It is easy to see that for propagation along $B$ the vacuum behaves
as in the limit $B=0$ for all eigenmodes. For that reason, we are
mainly interested in studying perpendicular photon propagation case
$k_{\parallel}=0$, for which the first mode is non physical. In
\cite{EliPRD} we solved numerically the system of equations
(\ref{FAMM2}) and (\ref{dispeq}), in the interval $0<B<B_c$ and
confirmed that the paramagnetic behavior is maintained throughout
the region of transparency. We  stress here that the photon magnetic
moment has  a maximum on the photon dispersion curve \cite{EliPRD}
near the threshold for pair creation $z_1 = -4m^2 + \epsilon$.

\subsection{The limit $k^2=z_1 +z_2<<eB$}

 There is wide range of frequencies characterized by the
condition $k^2=z_1 +z_2<<eB$, which corresponds to small deviations
from the light cone $k^2=0$. For such frequencies the photon
magnetic moment behavior is well described by the following
approximate expression (see the Appendix for details)
\begin{eqnarray}\label{FAMM2} \nonumber
   \mu_{\gamma}&=&-\frac{\partial \omega}{\partial B} \\
   &=& \displaystyle \frac{1}{2 \omega}\frac{\frac{\partial \chi_{i}^{(0)}}{\partial
  B} +\frac{\partial \chi_{i}^{(1)}}{\partial
  B}k^{2}+\frac{\partial \chi_{i}^{(2)}}{\partial
  B}k^{4}+\frac{\partial \chi_{i}^{(3)}}{\partial
  B}k^{6}}{1-\chi_{i}^{(1)}-\frac{\partial \chi_{i}^{(0)}}{\partial
  z_1}-Xk^{2}-Yk^{4}-\frac{\partial \chi_{i}^{(3)}}{\partial
  z_1 }k^{6}},\\ \nonumber
  X&=&\frac{\partial \chi_{i}^{(1)}}{\partial
  z_1}+2 \chi_{i}^{(2)},\\ \nonumber
  Y&=&\frac{\partial \chi_{i}^{(2)}}{\partial
  z_1}+3 \chi_{i}^{(3)},
   \end{eqnarray}
where the $\chi_{i}^{(l)}$ are functions of $z_1$ and $B$
\begin{eqnarray}
 \chi_{i}^{(l)} &=&\frac{2\alpha}{\pi} \int_{0}^{\infty}dt
\int_{-1}^{1}d\eta e^{-\frac{t}{b} }\psi_i^l , \\
\nonumber
  \psi_{i}^{(0)} &=& \frac{\rho_{0i}}{\sinh
t}e^{\zeta_{0}} , \\ \label{chi} \nonumber
 \psi_{i}^{(1)}
&=& \left[\frac{\rho_{0i}\xi+
\theta_i}{\sinh t}e^{\zeta_{0}} +\frac{1-\eta ^2}{8t} \right], \\
\nonumber \psi_{i}^{(l)} &=& \frac{e^{\zeta_{0}}}{\sinh
t}\left[\frac{\rho_{0i}\xi^{l}}{l!}+
\frac{\theta_i\xi^{l-1}}{(l-1)!}   \right], \hspace{1cm} l=2,3,... ,
\end{eqnarray}
and $\omega$, $k^{2}$ are given by
\begin{eqnarray}\nonumber
   k^2&=&\frac{\sqrt[3]{R
   +\sqrt{\frac{Q^3}{\left(\chi_{i}^{(3)}\right)^2}+R}}+\sqrt[3]{R
   -\sqrt{\frac{Q^3}{\left(\chi_{i}^{(3)}\right)^2}+R}}}{\sqrt[3]{\left(\chi_{i}^{(3)}\right)^{2}}}\\
\label{sol} & &\hspace{5.6cm}-\frac{\chi_{i}^{(2)}}{3\chi_{i}^{(3)}},\\
\nonumber
 R &=& -\frac{1}{2}\left[\chi_{i}^{(0)}\chi_{i}^{(3)}-
   \frac{1}{3}\left(\chi_{i}^{(1)}-1\right)\chi_{i}^{(2)}+\frac{2}{27}\left(\chi_{i}^{(2)}\right)^{2}\right],
   \\ \nonumber
  Q &=& \frac{1}{3}\left[
  \left(\chi_{i}^{(1)}-1\right)\chi_{i}^{(3)}-\frac{1}{3}\left(\chi_{i}^{(2)}\right)^{2}\right].
\end{eqnarray}
\begin{figure}
\centering
\includegraphics[width=0.55\textwidth]{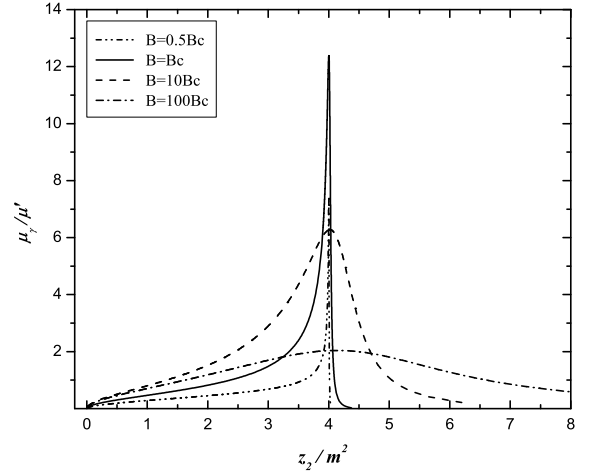}
\caption{ Photon anomalous magnetic moment for the second mode. }
\label{fig:MM}
\end{figure}
We have found a cubic-in-$k^2$ approximation for the dispersion
curve, which as we see in Fig. \ref{fig:mb2}, is valid in the whole
region of transparency for  small deviations from the light cone
dispersion equation.  The approximate curves can be compared to the
exact ones, which in both cases were drawn numerically. We conclude
from Fig. \ref{fig:MM} that in the region of transparency the
paramagnetic photon behavior is maintained for supercritical fields:
by fixing $z_2$, we observe that the quantity $-z_1$, and in
consequence $\omega$, decreases with increasing $B$.

\section{The decrease of frequency with increasing field}
In \cite{EliPRD} we showed that in the region of transparency
$\mu_{\gamma}^{i}=-\partial \omega^{i}/\partial B>0$, which means
that $\partial \omega^{i}/\partial B<0$. This means that the
frequency decreases with increasing field, that is, the incoming
photon is red-shifted. (In the case of a gravitational field, for
the incoming photon  the frequency increases with increasing the
modulus of the field \cite{gravredshift}). We will give below
detailed expressions, especially for the small departure from the
light cone case.

It is very important to consider at this point two limits for the
dispersion equations: the low frequency quasi-photon limit $\omega
\ll 2m$ (small departure from the light cone), and the high
frequency quasi-pair limit, which occurs for the second mode when
$z_1 \lesssim -4m^2$. In this case $\kappa^{2}$ has a inverse square
root divergence, and the solution of the dispersion equation shows a
very strong departure from the light cone. In the first case, the
expansion of $\kappa^{i}$ in the low frequency, low magnetic field
$b=B/B_c < 1$ limit, and the resulting dispersion equations, was
discussed in \cite{EliPRD}. The dispersion equation, written as
$\omega^2=|{\bf{k}}^2| + \kappa^{(i)}(z_2, eB)$ is such that
$\kappa^{(i)}(z_2, eB)\ll \omega^2$. Thus, as said earlier, the
photon self-energy acts as a small perturbation to the light cone
equation. The high frequency limit was discussed in \cite{shabad2}.
In that case, for instance, for the second mode, near the first
resonance frequency $z_1\lesssim -4m^2$, it is $\kappa^{(2)}(z_2,
eB)\gg \omega^2$, (in other words, for Landau quantum numbers $n,
n^{\prime}=0,1,2,.. $ the polarization tensor has an infinite set of
branching points at values $z_1= (E_{0n} +E_{0n^{\prime}})^2 $,
where $E_{0n}=\sqrt{m^2 +2eBn}$ and $E_{0n^{\prime}}=\sqrt{m^2
+2eBn^{\prime}}$). The polarization operator diverges and it is not
strictly a ``perturbation" but becomes the dominant term. It leads
to a quasi-particle which behaves as a massive vector boson, and we
name it quasi-pair. Its phase and group velocities are smaller than
$c$.

For the quasi-photon limit, by taking the first two terms in the
$\kappa^{(i)(0)}$ series expansion in powers of $b^2$, and up to
fields very close to $B_c$ (for instance, $B\sim 0.4 B_c$), in the
series expression for the functions $f^{(2,3)}$, defined in Sec. 2,
one can neglect terms from the power $b^4$ on. One has
$f^{(2,3)}/k_{\perp}^2 =-\frac{C^{i} \alpha b^2}{45 \pi }\ll 1$, and
as a good approximation the dispersion equations for these modes we
have,
\begin{equation}
\omega^{i2}-k^2_{\parallel}=k_{\perp}^2 \left(1- \frac{C^{i} \alpha
b^2}{45 \pi }\right) \label{Dispeq}
\end{equation}
where $C^{i}=7,4$ for $i=2,3$.  Eq. (\ref{Dispeq}) must be
interpreted as the dispersion equation in presence of the magnetic
field for an incoming photon which initially, far from the
magnetized region, satisfied the usual light cone equation
$\omega_0^2=k^2_{\parallel} + k_{\perp}^2$.  In other words, the
dispersion equation before the magnetic field was switched on. The
effect of the magnetic field is to decrease the incoming transverse
momentum squared by a factor $g(B)^{(i)}=1- f(B)^{(i)}/k_{\perp}^2
<1$ , to the effective value $k_{eff \perp}^2=k_{\perp}^2 g(B)^{(i)}
<k_{\perp}^2$ (and in consequence, the initial photon energy
decreased from $\omega_0 \to \omega$, where $\omega
=\sqrt{k^2_{\parallel}+k_{eff \perp}^2 }$. Thus, as stated
previously, the transverse momentum is not conserved in the magnetic
field, and $k_{eff \perp}$ is the effective transverse momentum
measured by an observer located in the region where the magnetic
field is $\mathbf{B}$. For propagation orthogonal to $\mathbf{B}$,
it is $\omega= \omega_0\sqrt{g(B)^{(i)}} $, since
$\omega_0=k_{\perp}$. The non conservation of momentum leads to the
decrease of the photon energy, which is red-shifted for incoming
photons.

 The magnetic field drains (gives) momentum (and
energy) to the incoming (outgoing) photon. The case is just the
opposite of the gravitational case, in which the gravitational field
increases (decreases) the incoming (outgoing) photon momentum (and
energy).

 Let us devote some space to remind the gravitational field case (we shall
 use in this paragraph the speed of light as $c$). The last statements can be seen by starting from the
  Hamilton-Jacobi equation in the  massless  limit (the action function $S$ becomes the eikonal).
\cite{Landau}.   For  a photon moving in  a centrally symmetric
gravitational field the constants of motion are the energy
$\omega_0$ and angular momentum $L$ with regard to its center. The
linear momentum is not a constant of motion.  Very far from the
massive body, the total energy is $\omega_0$, its linear momentum is
$k_0= \omega_0/c$.  Near the massive body of mass $M$,  for $r>r_G$,
by calling $e^{\nu}= 1- r_G/r$, where $r_G=2GM/c^2$ is the
gravitational radius of the body, we can write for a massless
particle whose squared effective radial momentum defined by $k_r^2=
e^{\nu}(\partial S/\partial r)^2$ as
\begin{equation}
k_r^2 + \frac{L^2}{r^2}= e^{-\nu}k_0^2 ,
\end{equation}
which expresses the total effective squared momentum as the
effective squared energy $\omega_G^2 =e^{-\nu}\omega_0^2$ divided by
$c^2$. The observed photon energy (frequency) has been increased
from $\omega_0$ to $\omega_G= e^{-\nu/2}\omega_0$.  Notice that for
$r_G \ll r$, one may write, by taking approximately $\omega_G \simeq
(1+ r_G/2r) \omega_0$,
\begin{equation}
c\sqrt{k_r^2 + (L^2/r^2)} - \frac{r_G \omega_0}{2r} = \omega_0
\label{conservE}
\end{equation}
which expresses in a transparent way the energy conservation, and
that the  observed (kinetic) energy for the approaching photon is
$\omega_G>\omega_0$ \cite{Landau}, whereas its interaction energy
with the body of mass $M$ is negative. For very large $r$,
(\ref{conservE}) leads back to $k_0 c = \omega_0$.

\subsection{Speed of light orthogonal to $\mathbf{B}$ and
vacuum compression} Lorentz transformations in non-parallel
directions change the magnetic field to $\mathbf{B}^{\prime}$ and
leads to the arising of an electric field  $\textbf{E}^{\prime}$,
preserving the invariance of ${\cal F}=2B^2 =2(B^{\prime 2} -
E^{\prime 2})$, but leading to  inequivalent solutions of the
equations of motion. However, they are physically good. Lorentz
frames parallel to $\mathbf{B}$ are preferred to preserve the
simplicity of the case $\mathbf{B}\neq 0$, $\mathbf{E}=0$. In all of
them the photon propagation have equivalent dynamics. It is easy to
see that $\partial\omega/\partial k_{\parallel}=1$ in these frames.

As the transverse momentum is not conserved, the speed of light
orthogonal to $\mathbf{B}$, if taken as
\begin{equation}
\partial \omega^i/\partial
k_{\perp}<1 \label{nonc}
\end{equation}
seems to lead to a sub-luminal speed of photons. This
interpretation, however, is logically unsatisfactory: one starts
from a relativistic invariant theory (Quantum Electrodynamics) and
from results obtained  perturbatively in the context of this theory
in a magnetized medium, concludes that the cornerstone of the
relativistic invariance is violated. We maintain the relativistic
principle of constancy of the speed of light in vacuum as valid, and
claim that (\ref{nonc}) expresses the fact that the non-conserved
momentum transverse to the field $\textbf{B}$ has an effective value
smaller than $k_{\perp}$. In doing that, we state that due to the
non-conservation of transverse momentum $k_{\perp}$, both its
initial value $k_{\perp}$ and energy $\omega_0$ have been decreased
to  $k_{eff\perp}$, $\omega^i$ and the transverse speed of light
must be expressed by the equation $\partial \omega^i/\partial k_{eff
\perp}=1$, in full analogy to the gravitational field case. That is,
local observers would find the transverse speed of light as unity.
For them, from (\ref{Dispeq}), the light cone equation can be
written in coordinate space as
\begin{equation}
\left[\frac{\partial^2 }{\partial x_1^{2i\prime}} + \frac{\partial^2
}{\partial x_2^{2i\prime}}+ \frac{\partial^2}{\partial
x_3^2}-\frac{\partial^2 }{\partial x_0^2}\right]\Psi^{i}=0
\end{equation}
where $x_{1,2}^{i\prime}=x_{1,2}^{i}/\sqrt{g(B)^{(i)}}>x_{1,2}^{i}$.
This means that the local observer measures, for instance, longer
wavelengths, since any rule for measuring lengths if placed in
magnetized vacuum,  is compressed in the direction orthogonal to
$\mathbf{B}$ in the amount $\sqrt{g(B)^{(i)}}$. The longer
wavelength is in correspondence to the observed smaller frequencies
$\omega^{i} < \omega_0$. The vacuum compression is due to  the
negative pressure effect of magnetized vacuum in the direction
perpendicular to the field $\mathbf{B}$ discussed in
\cite{Elizabeth}. Such compression is related to the following
facts:  the quantity $S_B=c\hbar/eB$ can be considered as the
quantum of area corresponding to a flux quantum for a field
intensity $\mathbf{B}$. Thus, by increasing $B$, $S_B$ decreases. As
a consequence, the spread of the electron and positron wave
functions decreases exponentially with $B$ in the direction
orthogonal to the field since they depend on the transverse
coordinates as $e^{-\xi^2}$ where $\xi^2=x_{\perp}^2/S_B$.

These results mean  space-time consequences which bear some analogy
to general relativity: we have seen that the vacuum orthogonal to
the field behaves as compressed; and also that the red shift means
shorter frequencies. But this, in turn, leads to the fact that if
time is measured by the wave modes periods $T^{(i)}=2
\pi/\omega^{(i)}$, it runs faster for increasing $B$ and do it in a
polarization-dependent way and for waves propagating non parallel to
$\bf{B}$.

The previous discussion is valid for the low frequency $\omega \ll
2m$, low magnetic field limit, $B \ll B_c$, when the spacing between
Landau levels is small  compared to $2m$. As the field intensity
increases the quantity $g(B)$ decreases. The role of the separation
between Landau levels of virtual pairs becomes more and more
significant as one approaches the first threshold of resonance,
which is the quasi-pair region, where $B \lesssim B_c$. For
frequencies $\omega \simeq 2m$ and $k_{\parallel}< \omega$, the
dispersion equation for the second mode may be written
\cite{shabad2}, since the polarization operator is expressed as a
sum over Landau levels $n, n^{\prime}$ of the virtual
electron-positron pairs, in terms of the dominant term $n
=n^{\prime}=0$, as
\begin{equation}\label{expDE}
z_1 + z_2 =\frac{2\alpha eB m e^{-z_2/2 eB}}{\sqrt{z_1 + 4m^2}}.
\end{equation}
This equation is valid in a neighborhood of $z_1 \lesssim -4m^2$.
Notice that its limit for $\bf{k} \to 0$ is $\omega \neq 0$.
Actually, it describes a massive vector boson particle closely
related to the electron-positron pair (see below). This is not in
contradiction with the gauge invariance property of the photon self
energy. Eq. (\ref{expDE}) has  solutions found by Shabad
\cite{shabad2} as those of a cubic equation. One can estimate its
behavior very near $z_1 =-4m^2$, by assuming $z_1 =-4m^2 + \epsilon$
and $z_2 =4m^2 - \epsilon$, the initial energy and transverse
momentum where $\epsilon$ is a small quantity. One can obtain the
solution approximately as $(z_1 + 4m^2)^{3/2}=2\alpha eB m e^{-z_2/2
eB}$, from which $z_1=-4m^2 +(2\alpha eB m e^{-z_2/2 eB})^{2/3} $.
This means approximately $\omega^2= \sqrt{k_{\parallel}^2 + 4m^2 -
(2\alpha eB m e^{-2m^2/ eB})^{2/3}}$.  Thus, the transverse momentum
of the original photon is trapped by the magnetized medium, the
resulting quasi-particle being deviated to move along the field as a
vector boson of mass $\omega_t= \sqrt{4m^2 -m^2(2\alpha b
e^{-2/b})^{2/3}}$. Our approach is approximate. A more complete
discussion would be made by following the method of \cite{shabad2}.
This quasi-pair is obviously paramagnetic, as can be checked easily.
It differs totally from photons originally propagating parallel to
$\mathbf{B}$. For slightly larger energies such that $z_1 \lesssim
-4m^2$, and $b$ of order unity, that is $B \sim B_c$, they decay in
observable electron-positron pairs, and the polarized vacuum becomes
absorptive (see \cite{shabad2}). Thus, near the critical field $B_c$
our problem bears some analogy to the gravitational singularity
effects on light. For light passing near a black hole, if $r  \simeq
r_G$, the light is deviated enough to be absorbed by the black hole.
Among other differences in both cases, it must be remarked that the
gravitational field in black holes is usually centrally symmetric,
whereas our magnetic field is axially symmetric.

\section{The red-shift of the paramagnetic photon}
For the specific case of the magnetic field produced by a star, we
assume that it has axial symmetry and that it decreases with
increasing distance along the plane orthogonal to it. In place of
assuming an explicit dependence $\mathbf{B}=\mathbf{B}(\textbf{r})$,
we assume a partition in concentric shells, in which the magnetic
field is considered as constant inside  each one. Then $B$ increases
to $B + \Delta B$ when passing from a shell to its inner neighbor,
and decreases $B - \Delta B$ when passing to the outer one.

From (\ref{Dispeq}), the frequency is red shifted  when passing from
a region of magnetic field $B$ to another of increased field $B +
\Delta B$. In the same limit it is,
\begin{equation}
\Delta \omega^{(2)}=- \frac{14 \alpha z_2 b \Delta b}{45 \pi
|\textbf{k}|}<0,\label{si}
\end{equation}
and

\begin{equation}
\Delta \omega^{(3)}= -\frac{8 \alpha z_2 b \Delta b}{45 \pi
|\textbf{k}|}<0. \label{M3}
\end{equation}

Here $\Delta b =\Delta B/B_c$. Thus, the red shift, consequence of
the photon paramagnetism, differs for longitudinal and transverse
polarizations.

To give an order of magnitude, for instance, for photons of
frequency $10^{20}$ Hz, and magnetic fields of order $10^{12}$ G,
$|\Delta \omega| \sim 10^{-6}\omega$.

For the quasi-pair case, from \cite{EliPRD}, when $z_1 \to -4m^2 +
\epsilon$ the photon redshift can be written approximately, by
calling $P=4m^2 + z_1$. $U=\alpha m^3 e^{-z_2/2eB}$, as
\begin{equation}
\Delta \omega^{(2)}= -\frac{PU}{\omega B_c (P^{3/2}+
bU)}\left(1+\frac{z_2}{2eB}\right)\Delta B<0,\label{FRR1}
\end{equation}

The coefficient of $\Delta B$ at the right, which is minus the
photon magnetic moment, has  a maximum located on the dispersion
curve near the threshold for pair creation $z_1 = -4m^2 + \epsilon$.
In terms of \begin{equation}\omega_t=\sqrt{4m^2 - m^2[2\alpha
b\exp{(-2/b)}]^{2/3}},\end{equation} this maximum is
\begin{equation}\label{FAMMaprox1}
\mu_{\gamma}^{(2)}= \frac{e(1+2b)}{3\omega_t b}\left[2\alpha
b\exp{(-2/b)}\right]^{2/3}
\end{equation}
\noindent which for $b \sim 1$ is about  $13 \mu^{\prime}$, where
$\mu^{\prime}$ is the anomalous electron magnetic moment.

For supercritical fields $B \to B_c/4 \alpha$, $\mu_{\gamma}^{(2)}$
(given by the expression (\ref{FAMMaprox1})) may become arbitrarily
large. But this formula is not valid in the mentioned limit $B \to
B_c/4 \alpha$, for which  the condition $z_1\ll4m^2$ is also
satisfied when $z_2\approx4m^2$. In that case, according to
\cite{Shabad3}, the right approximate dispersion equation is
\begin{equation}\label{dispeqaprox2}
    z_1=-\frac{z_2}{1+\frac{\alpha}{3\pi}b},
\end{equation}
and, as a consequence,  the photon anomalous magnetic moment (for
perpendicular photon propagation)  looks like
\begin{equation}\label{FAMMaprox2}
\mu_{\gamma}^{(2)}= \frac{\alpha e}{6\pi
m^2}\frac{\sqrt{z_2}}{1+\frac{\alpha}{3\pi}b}.
\end{equation}
It is easy to see from (\ref{FAMMaprox2}) that $\mu_{\gamma}^{(2)}$
uniformly tends to zero when the magnetic field grows
$b\rightarrow\infty$.

 Notice that
(\ref{si}),(\ref{M3}) are the analog of the gravitational red shift
\cite{gravredshift}
\begin{equation}
\Delta \omega_g = -\frac{ \Delta \phi}{c^2}\omega,
\end{equation}
where $\Delta \phi= -GM/r_2 + GM/r_1$ and $r_2>r_1$. However since
the gravitational field is negative, $\Delta \phi>0$ corresponds to
a decrease in the absolute value of $\phi$, as opposite to $\Delta B
> 0$. But as pointed out earlier, the magnetic red shift is produced with opposite sign than the gravitational red shift.
For $r_2 \to \infty$ the photon gravitational red shift is $\Delta
\omega_g = -r_G/2r$; this is  what is observed for the light coming
from a  star of mass $M$. For a neutron star of mass $M \sim
M_{\bigodot}$, and star radius  $r_1\sim 10$ Km, $\omega_g/\omega
\sim 10^{-1}$. The magnetic red shift for the same star, at
frequencies of order $\omega=2m$ and field $B \sim B_c$ is $\Delta
\omega_B/\omega = \int_0^B \mu_{\gamma}^{i}dB/\omega \sim 10^{-5}$.
This implies that the magnetic red shift is a small correction to
the gravitational red shift up to critical fields.

\section{Conclusions}
We have shown that  the photon magnetic moment
${\mathbf{\mu}}_{\gamma}^{(i)}$ can be understood as as a
pseudovector quantity, which is linear in the electromagnetic field
tensor ${\cal{F}_{\mu \nu}}$.  A cubic in $k^2$ approximation for
the polarization operator was obtained, from which analytic
solutions of the photon dispersion equations and anomalous magnetic
moment are easily deduced. These approximate expressions are valid
in a very wide range of photon momentum and magnetic fields,
whenever the condition $k^2/eB\ll 1$ is satisfied.  In the whole
region of transparency the paramagnetic photon behavior is
maintained, even for supercritical fields $B > B_c$.

  In the region of transparency  and for magnetic fields $B\ll B_c$ and frequencies $\omega \ll 2m$
photons propagate in magnetized vacuum with energies and transverse
momentum decreasing for increasing fields, and vice-versa: it
behaves as a tiny dipole moving at the speed of light in magnetized
vacuum.  For larger magnetic fields $B\simeq B_c$ and frequencies
$\omega \lesssim 2m$, the resulting quasi--particle behaves as a
massive  vector boson moving parallel to the field \textbf{B}, its
mass being $m_q \lesssim 2m$. The last behavior extends to all the
region of transparency for supercritical fields $B\gg B_c$. It has
been discussed the analogy between the photon propagation in a
magnetic field and in a gravitational field. Red shift is produced
also in the magnetic field case, but with opposite sign than the
gravitational one, leading also to space-time deformations.

The presented  results, related to the photon propagation in a
uniform magnetic field, may be applied to the study of photons in an
axially symmetric magnetic field $B= B(x_{\perp})$, by considering
concentric shells in which the field is taken as uniform, but
varying from shell to shell. This can be made whenever the variation
of $B$ over the length $l= \sqrt{\hbar c/eB} $ is negligibly small.
We have found a  that the photon magnetic moment has a maximum on
the dispersion curve, in the region close to the electron-positron
pair creation threshold.

The study of photon properties in an external magnetic field
\cite{Adler}-\cite{ShabadUsov} is
 very important in the astrophysical
context, where  high magnetic  fields  have been estimated to exist
\cite{GRB}-\cite{magnAstroph2}; and can be considered as an
important part of a more general problem: the theoretical study of
high energy processes of elementary particles
  in strong external electromagnetic fields.  Nowadays, this issue has also
attracted the interest of several experimental researchers,
   due to  the development of high
power lasers and ion accelerators (see \cite{tev},\cite{Muller} and
references therein).

\begin{acknowledgements}
The authors thank A.E. Shabad for some comments, and especially to
OEA-ICTP for support
  under Net-35. E.R.Q. also
thanks ICTP for hospitality.
\end{acknowledgements}

\appendix
\section{Series expansion of the polarization operator}
\subsection{$k^2=z_1 +z_2<<eB$ limit}
The functions $\zeta$ and  $\rho_i$ ($i=1,2,3$), linear in $z_1$ and
$z_2$, may also be written as
\begin{eqnarray}
\zeta(z_1,-z_1+k^2,B)&=&\zeta_{0}(z_1,B)+k^2\xi(B),\\
 \rho_i(z_1,-z_1+k^2)&=& \rho_{0i}(z_1)+k^2  \theta_i,
\end{eqnarray}
where  $\zeta_{0}=\zeta(z_1,-z_1,B)$, $ \xi=\zeta(0,k^2,B)/k^2$ and
$\rho_{0i}=\rho_{i}(z_1,-z_1)$, $\theta_i=\rho_{i}(0,k^2)/k^2$. We
can express then (\ref{op-pol2}) as
\begin{equation}\label{series-op-pol}
\kappa_{i}=\sum_{l=0}^{\infty} \chi_{i}^{(l)}k^{2l},
\end{equation}
\begin{eqnarray}
 \chi_{i}^{(l)} &=&\frac{2\alpha}{\pi} \int_{0}^{\infty}dt
\int_{-1}^{1}d\eta e^{-\frac{t}{b} }\psi_i^l , \\
\nonumber
  \psi_{i}^{(0)} &=& \frac{\rho_{0i}}{\sinh
t}e^{\zeta_{0}} , \\ \nonumber
 \psi_{i}^{(1)}
&=& \left[\frac{\rho_{0i}\xi+
\theta_i}{\sinh t}e^{\zeta_{0}} +\frac{1-\eta ^2}{8t} \right], \\
\nonumber \psi_{i}^{(l)} &=& \frac{e^{\zeta_{0}}}{\sinh
t}\left[\frac{\rho_{0i}\xi^{l}}{l!}+
\frac{\theta_i\xi^{l-1}}{(l-1)!}   \right], \hspace{1cm} l=2,3,... .
\end{eqnarray}

We retain  only the first four terms  in the series expansion
(\ref{series-op-pol})
\begin{equation}\label{PolOp}
 \kappa_{i}\approx
\chi_{i}^{(0)}+\chi_{i}^{(1)}k^{2}+\chi_{i}^{(2)}k^{4}+\chi_{i}^{(3)}k^{6},
\end{equation} and explicitly solve the
resulting approximate dispersion equation, cubic-in-$k^2$,
\begin{equation}\label{DE}
 0=
\chi_{i}^{(0)}+(\chi_{i}^{(1)}-1)k^{2}+\chi_{i}^{(2)}k^{4}+\chi_{i}^{(3)}k^{6}.
\end{equation}
The solutions are given by
\begin{eqnarray}\nonumber
   k^2&=&\frac{\sqrt[3]{R
   +\sqrt{\frac{Q^3}{\left(\chi_{i}^{(3)}\right)^2}+R}}+\sqrt[3]{R
   -\sqrt{\frac{Q^3}{\left(\chi_{i}^{(3)}\right)^2}+R}}}{\sqrt[3]{\left(\chi_{i}^{(3)}\right)^{2}}}\\\label{sol}
 & &\hspace{5.2cm}-\frac{\chi_{i}^{(2)}}{3\chi_{i}^{(3)}},\\
\nonumber
 R &=& -\frac{1}{2}\left[\chi_{i}^{(0)}\chi_{i}^{(3)}-
   \frac{1}{3}\left(\chi_{i}^{(1)}-1\right)\chi_{i}^{(2)}+\frac{2}{27}\left(\chi_{i}^{(2)}\right)^{2}\right],
   \\ \nonumber
  Q &=& \frac{1}{3}\left[
  \left(\chi_{i}^{(1)}-1\right)\chi_{i}^{(3)}-\frac{1}{3}\left(\chi_{i}^{(2)}\right)^{2}\right].
\end{eqnarray}

We differentiate with regard to $B$ the dispersion equation
(\ref{DE})   and, by using  $\frac{\partial z_{1}}{\partial
B}=-2\omega \frac{\partial \omega}{\partial B}$ ,  we obtain finally
an expression for the photon anomalous magnetic moment
\begin{eqnarray}\label{FAMM2} \nonumber
   \mu_{\gamma}&=&-\frac{\partial \omega}{\partial B} \\
   &=& \displaystyle \frac{1}{2 \omega}\frac{\frac{\partial \chi_{i}^{(0)}}{\partial
  B} +\frac{\partial \chi_{i}^{(1)}}{\partial
  B}k^{2}+\frac{\partial \chi_{i}^{(2)}}{\partial
  B}k^{4}+\frac{\partial \chi_{i}^{(3)}}{\partial
  B}k^{6}}{1-\chi_{i}^{(1)}-\frac{\partial \chi_{i}^{(0)}}{\partial
  z_1}-Xk^{2}-Yk^{4}-\frac{\partial \chi_{i}^{(3)}}{\partial
  z_1 }k^{6}},\\ \nonumber
  X&=&\frac{\partial \chi_{i}^{(1)}}{\partial
  z_1}+2 \chi_{i}^{(2)},\\ \nonumber
  Y&=&\frac{\partial \chi_{i}^{(2)}}{\partial
  z_1}+3 \chi_{i}^{(3)},
   \end{eqnarray}
   with $\omega$ and $k^{2}$ given by (\ref{sol}) and
\begin{eqnarray} \frac{\partial \chi_{i}^{(l)}}{\partial
  B}&=&-\frac{2\alpha}{\pi B} \int_{0}^{\infty}dt
\int_{-1}^{1}d\eta e^{-\frac{t}{b} } \frac{e^{\zeta_{0}}}{\sinh
t}\vartheta_{i}^{(l)}\\ \nonumber
 \vartheta_{i}^{(0)}&=& \rho_{0i}\left[-\frac{t}{b}+\zeta_{0}\right] , \\ \nonumber
\vartheta_{i}^{(1)} &=& \left[\left(\rho_{0i}\xi+
\theta_i \right)\left(-\frac{t}{b}+\zeta_{0}\right)
+\rho_{0i}\xi\right]-\frac{\sinh
t}{e^{\zeta_{0}}}\frac{1-\eta ^2}{8b}, \\
\nonumber \vartheta_{i}^{(2)}&=&
\left[\left(\frac{\rho_{0i}\xi^{2}}{2}+
\theta_{i}\xi\right)\left(-\frac{t}{b}+\zeta_{0}\right)
+\rho_{0i}\xi^{2}+ \theta_{i}\xi
\right],\\
\nonumber \vartheta_{i}^{(3)}&=&
\left[\left(\frac{\rho_{0i}\xi^{3}}{6}+
\theta_{i}\frac{\xi^2}{2}\right)\left(-\frac{t}{b}+\zeta_{0}\right)
+\rho_{0i}\frac{\xi^{3}}{3}+
\theta_{i}\xi^2 \right],
\end{eqnarray}
\begin{eqnarray} \frac{\partial \chi_{i}^{(l)}}{\partial
  z_1}&=&-\frac{2\alpha}{\pi z_1} \int_{0}^{\infty}dt
\int_{-1}^{1}d\eta e^{-\frac{t}{b} }\upsilon_{i}^{(l)}\\ \nonumber
 \upsilon_{i}^{(0)}&=&\frac{\rho_{0i}}{\sinh
t}e^{\zeta_{0}}\left(\zeta_{0}+1\right) , \\ \nonumber
\upsilon_{i}^{(1)}&=& \frac{e^{\zeta_{0}}}{\sinh
t}\left[\left(\rho_{0i}\xi+
\theta_i \right)\zeta_{0}+\rho_{0i}\xi\right], \\
\nonumber \upsilon_{i}^{(2)}&=& \frac{e^{\zeta_{0}}}{\sinh
t}\left[\left(\frac{\rho_{0i}\xi^{2}}{2}+
\theta_{i}\xi\right)\zeta_{0}+\rho_{0i}\frac{\xi^{2}}{2}
\right],\\
\nonumber \upsilon_{i}^{(3)}&=& \frac{e^{\zeta_{0}}}{\sinh
t}\left[\left(\frac{\rho_{0i}\xi^{3}}{6}+
\theta_{i}\frac{\xi^2}{2}\right)\zeta_{0}+\rho_{0i}\frac{\xi^{3}}{6}
\right].
\end{eqnarray}




\end{document}